\pdfoutput=1 
\documentclass[sigconf]{acmart}
\usepackage{booktabs} 
\usepackage{amsmath}
\usepackage{tikz-cd} 

\setcopyright{rightsretained}

\acmConference[FAT/ML 2017]{FAT/ML 2017}{August 2017}{Halifax, Nova Scotia, Canada} 
\acmYear{2017}
\copyrightyear{2017}

\begin{document}
\title{Fair Pipelines}
\titlenote{Presented as a poster at the 2017 Workshop on Fairness, Accountability, and Transparency in Machine Learning (FAT/ML 2017)}

\author{Amanda Bower}
\affiliation{%
\institution{University of Michigan}
  }
\email{amandarg@umich.edu}

\author{Sarah N. Kitchen}
\affiliation{%
\institution{Michigan Tech Research Institute}
}
\email{snkitche@mtu.edu}

\author{Laura Niss}
\affiliation{%
  \institution{University of Michigan}
}
\email{lniss@umich.edu}

\author{Martin~J.~Strauss}
\affiliation{%
  \institution{University of Michigan}
}
\email{martinjs@umich.edu}

\author{Alexander Vargo}
\affiliation{University of Michigan
}
\email{ahsvargo@umich.edu}

\author{Suresh Venkatasubramanian}
\affiliation{University of Utah
}
\email{suresh@cs.utah.edu}
\begin{abstract}
This work facilitates ensuring fairness of machine learning in the real world by decoupling fairness considerations in compound decisions. In particular, this work studies how fairness propagates through a compound decision-making processes, which we call a \textit{pipeline.} Prior work in algorithmic fairness only focuses on fairness with respect to one decision. However, many decision-making processes require more than one decision. For instance, hiring is at least a two stage model: deciding who to interview from the applicant pool and then deciding who to hire from the interview pool. Perhaps surprisingly, we show that the composition of fair components may not guarantee a fair pipeline under a $(1+\varepsilon)$-equal opportunity definition of fair. However, we identify circumstances that do provide that guarantee. We also propose numerous directions for future work on more general compound machine learning decisions.


\end{abstract}

%
%





\maketitle
\renewcommand{\shortauthors}{A. Bower, S. Kitchen, L. Niss, M. Strauss, A. Vargo, S. Venkatasubramanian}

\section{Introduction}

Automated-decision making saves time and is implicitly assumed to prevent human bias. However, such automated decisions may unfortunately lead to unfair outcomes. Until recently, the use of automated-decision making has been largely unchecked. In pursuit of this goal, the first question that needs to be addressed is what ``fairness" itself actually means and how to quantify it. A meta-analysis of the current literature indicates there are a multitude of inequivalent applications of the term, and consequently metrics. See, for example, \cite{fsv} (which is based off of definitions in the seminal paper, \cite{dhprz}) for a general framework that encompasses notions of individual fairness and group fairness (``non-discrimination'').

In this paper, we are particularly interested in how effects of bias compound in decision-making pipelines. While prior work in algorithmic fairness has focused on fairness of one decision, it is not immediately clear how fairness propagates throughout a compound decision making process. Complicated decisions usually require more than one decision. For example, a hiring process may include two decisions: from an applicant pool, one first decides who gets an interview, and the final hiring decision is made from the pool of interviewees. Although this is a relatively simple two-decision example, one can imagine a recursive-like compound decision-process where the outcome of one decision affects another and vice versa. For instance, perhaps to be brought in for an interview at company A, working for company B helps greatly, but working for company A also helps an applicant greatly to get an interview at company B. 

We ask the following questions:
\begin{enumerate}
\item Can a decision at point $j$ in a decision pipeline correct for unfairness at point $i<j$?
\item How much fairness from point $i$ is preserved in later points in the pipeline?
\item More specifically, how does the fairness from each stage contribute to the fairness of the final decision?
\end{enumerate}


Our contribution to the algorithmic fairness field is to highlight the need to study compound decision making processes by studying how composability and fairness interact. We emphasize that pipelines are useful to study because they decouple the intermediate decisions since there may be completely different parties with varying goals and mechanisms responsible for each decision. Perhaps suprisingly, even in the most basic example of a two-stage pipeline, we show under a $(1+\varepsilon)$-equal opportunity definition of fair, the two stages cannot necessarily be combined as expected. Finally, pipelines set the stage for a number of interesting questions detailed in Section \ref{futurework}.
 
 \section{Framework}
 
 \subsection{Pipelines}
\begin{definition}[Straight Pipeline] An $n$-stage (straight) pipeline $P(f,g)$ on a set $O$  is an ordered set of decision functions $$\mathcal{F}=\{f_1: O \rightarrow D_1, \ f_2: O \times \widehat{D_1} \rightarrow D_2, \ \dots, \ f_T: O \times \widehat{D_{T-1}} \rightarrow D_T \}$$ where $\widehat{D_t}= D_{k_t} \times D_{k_t+1} \times \cdots \times D_{t-1} \times D_{t}$ for $1 \leq k_t \leq t$ for $t = 1, \dots, T$ and rule functions $$\{g_1: D_1 \rightarrow \{0,1\}, \dots ,g_{T-1}: D_{T-1} \rightarrow \{0,1\}\}$$  where the final decision for $x \in O$ is given by \[   P(f,g)(x):=\left\{
\begin{array}{ll}
      \hat{f_T}(x) & g_t(\hat{f_t}(x))=1 \ \forall t \in [T-1] \\
      \mathrm{FAIL} & \mathrm{otherwise}
\end{array} 
\right. \]
where for $x \in O$, $\widehat{f_1}(x)= f_1(x)$ and for $t=2,\dots,T$, $\hat{f_t}(x) =f_t(x,\widehat{f_{k_t}}(x), \widehat{f_{k_t +1}}(x), \dots, \widehat{f_{t}}(x)).$
We will say decision function $f_t$ takes place at \textit{stage $t$} of the pipeline for $t=1, \dots, T$. 
\end{definition}

To understand the above definition, note that a straight pipeline $P(f,g)$ on a set $O$ (for instance, applicants to a job) takes input $x \in O$ and applies a decision function on $x$. If the first decision ($f_1(x)$), on $x$ is satisfactory (where satisfactory is determined by $g_1$), $x$ is passed onto the next decision function and so on and so forth until there are no more decisions to be made (reaching $f_T$) or an intermediate stage declares an unsatisfactory decision (determined by some $g_t$) at which point no further decisions shall be made. Each subsequent decision function after the first may see some part of the past decisions prior in the pipeline (how many decisions back decision function $f_t$ can see back is determined by $k_t$).

We expect the following variations and restrictions to be common with illustrative examples below:
\begin{enumerate}
\item \textbf{Filtering pipeline.} When each decision function $f_t$ is binary, take $g_t(0) = 0$ and $g_t(1) = 1$, i.e., only the positive-decision subset of previous stage gets passed on. In this case, we may use the notation $P(f,g)$ in place of $P_f$ for concision.
\item \textbf{Cumulative decisions pipeline.} Take $k_t \leq i$, i.e., the decision at each stage agglomerates some score onto previous stages' decision scores, so that each stage's decision can depend on some of the previous decisions.
\item \textbf{Informed pipeline.} Each stage of the pipeline has summary statistics about the outcome of previous stages.  In this paper, those statistics are implicit in the definition of $f_t$.  In particular, our notation above, while sufficient for this paper, is insufficient to study linked decisions in which each decision reacts to statistics about the other. 
\end{enumerate}
The below diagram illustrates a two-stage pipeline, where the second decision can see what happened in the first decision on a given input:

\noindent \begin{tikzcd}
 x \arrow[r, "f_1"]
  & f_1(x) \arrow[d,"g_1(f_1(x))=0"] \arrow[r,"g_1(f_1(x))=1"]
 & (x,f_1(x)) \arrow[r, "f_2"] & f_2(x,f_1(x)) \arrow[d]  \\
                                  &   P(f,g)(x) = FAIL && P(f,g)(x) = \hat{f}_2(x)
  \end{tikzcd}

\begin{example}[Hiring decisions as a two-stage filtering pipeline]
Let $O$ be the applicant pool, $f_1: O\to D_1=\{0,1\}$ the decision as to who receives an interview, and $f_2: O\times D_1 \to D_2=\{0,1\}$ the hiring decision.  Then $P_f(x) = \hat{f}_2(x)=f_2(x,f_1(x))$ for all applicants $x$ such that $f_1(x)=1$ (i.e. for all applicants who receive an interview), and $0$ (or FAIL) otherwise. 
\end{example}

\subsection{Fairness}
As we have stated, there is no consensus in the literature on the definition of fairness. However, there have been many  recent proposed definitions. For simplicity, in order to illustrate how fairness propagates through a filtering pipeline, we will build on the definition of equal opportunity found in \cite{hps}.



\subsubsection*{Equal Opportunity} 
We consider the case of making a binary decision $\hat{Y} \in \{0,1\}$ and measure fairness with respect to a protected attribute $A \in \{0, 1\}$ (such as age, gender, or race) and the true target outcome $Y\in \{0, 1\}$, which captures if an individual is qualified or not.
\begin{definition} As defined in \cite{hps}, a binary predictor $\hat{Y}$ satisfies equal opportunity with respect to $A$ and $Y$ if
	\begin{equation*}
	\Pr\{ \hat{Y}=1 | A = 0, Y = 1\} = 	\Pr\{ \hat{Y}=1 | A = 1, Y = 1\}.
	\end{equation*}
\end{definition}

	In other words, we make sure that the true positive rates are the same across a protected attribute. We usually think of $A=0$ as a majority class. 
    \subsubsection*{\textrm{$(1+\varepsilon)$}-Equal Opportunity}
Equal opportunity may be far too restrictive since it requires exact equality of two probabilities. In addition, because our goal is to measure how fairness propagates through a pipeline, we propose to quantify fairness relative to a majority class with an $\varepsilon$ factor. In fact, we propose a framework that consists of boosting the minority class in order to correct for existing bias. Therefore, we introduce the notion of ($1+\varepsilon$)-equal opportunity, which allows for compensation of inherent biases in training data. 

\begin{definition}[\textrm{$(1+\varepsilon)$}-equal opportunity] A binary predictor $\hat{Y}$ satisfies ($1+\varepsilon$)-equal opportunity with respect to $A$, $Y$, and majority class $A=0$ if

	\begin{equation*}
	(1+\varepsilon)\Pr\{ \hat{Y}=1 | A = 0, Y = 1\} \leq \Pr\{ \hat{Y}=1 | A = 1, Y = 1\},
	\end{equation*}
where $\varepsilon\in[0,1)$ can be any real number such that
\begin{equation*}
(1+\varepsilon)\Pr\{ \hat{Y}=1 | A = 0, Y = 1\} \in [0,1].
\end{equation*}
This definition generalizes to more than one protected class in a natural way: if $A=\{a_1, \dots, a_m\}$ and $a_m$ represents the majority class, then $\hat{Y}$ satisfies $((1+\varepsilon_1),\dots,(1+\varepsilon_m))$-equal opportunity with respect to $A$, $Y$, and $a_m$, if $$(1+\varepsilon_t)\Pr\{ \hat{Y}=1 | A = a_m, Y = 1\}\leq \Pr\{ \hat{Y}=1 | A = a_t, Y = 1\}$$ for $t=i1,\dots, m-1$.
\end{definition}

That is, we make sure that the true positive rates in the protected class are $1+\varepsilon$ times the rates in the majority class.  The factor of $(1 + \varepsilon)$ could be determined, for example, by a Human-Resources professional or lawyer in order to correct known bias, past or present, whose mechanisms may not be fully understood.  (The problem of choosing $\varepsilon$ properly is a much more difficult control problem, possibly involving feedback, and is deferred to later work.)

\subsection{Why pipelines?}



\subsubsection*{Examples}
We will now illustrate the utility of studying pipelines with a few examples.  Many more examples have been identified in \cite{o2016weapons}. 
\begin{enumerate}
\item \textbf{Hiring.} Hiring for a job is at least a two-stage pipeline: (1) determine who to interview out of an applicant pool and (2) determine who to hire out of an interview pool. Hiring is also an example of a \textit{filtering pipeline} since only those who have successfully got an interview are passed onto the interview stage. This pipeline can also be an example of a \textit{informed pipeline} if the final stage gets information about the racial, gender, and age make-up of the applicant pool for instance.
\item \textbf{Criminal Justice.} Getting parole can be thought of as a three-stage pipeline: (1) compute a defendant's risk assessment score, (2) if the defendant is convicted, determine the criminal's sentencing, and (3) determine whether the criminal gets parole. This pipeline is an example of a \textit{cumulative decisions pipeline} since the parole board has information about the risk assessment score and sentencing. 
\item \textbf{Mortgages} Getting a mortgage for a home can be thought of as a \textit{looping pipeline}. An applicant's FICO score is used to determine whether they get a mortgage. However, an applicant's FICO score is affected by prior credit and loan decisions, which also use the applicant's FICO score.
\end{enumerate}


In the following, we show that (under specific circumstances) the fairness of a compound process can be guaranteed by making each link in the pipeline fair.  This has the desirable implication that ``global" fairness can be obtained via ``local'' fairness under these specific circumstances.  In future work, we aim to develop a more general result that would guarantee fairness over the entire pipeline while allowing for each organization making one of the decisions in the pipeline to consider fairness only in its own decision.

\section{Results}
For the rest of the paper, we will focus on two-stage pipelines whose decision functions are binary decision functions, i.e., $D_1=D_2 = \{0,1\}.$ We will usually refer to such a decision function as a binary predictor $\hat{Y}$.



\subsection{Pipeline Fairness}
Consider a two-stage pipeline with binary predictors $f_1 = \hat{X}$ and $f_2=\hat{Y}$ where again we take $A=0$ to be the majority class and $X$ and $Y$ be the true target outcomes.  Using the hiring scenario from above, for example, $\hat{X}$ would represent the decision about whether or not to interview a  candidate and $\hat{Y}$ would represent the decision about whether or not to hire a candidate.  If $X =1$, then the candidate is qualified to get an interview; likewise, a candidate with  $Y=1$ is a good fit for the job.

We define the pipeline to be $(1+\alpha)$-equal opportunity fair if the final decision is $(1+\alpha)$-equal opportunity fair: $$(1+\alpha)\Pr\{\hat{Y}=1|Y=1, A=0\} \leq \Pr\{\hat{Y}= 1 |Y=1, A=1\}.$$ 
Assuming 
\begin{enumerate}
\item $(1+\varepsilon)\Pr\{\hat{X}=1|Y=1, A=0\} \leq \Pr\{\hat{X}=1|Y=1, A=1\}$, a nontrivial assumption that looks something like $(1+\varepsilon)$-equal opportunity for the first stage in the pipeline.
\\
\item $(1+\delta)\Pr\{\hat{Y}=1|\hat{X}=1, Y=1, A=0\}\leq \Pr\{\hat{Y}=1|\hat{X}=1, Y=1, A=1\},$ that is, $(1+\delta)$-equal opportunity for the second stage in the pipeline.\\
\item $\hat{Y}=1 \implies \hat{X}=1.$
\end{enumerate}
See Section 2.3 for a discussion of these assumptions.
Then, we have that 

\begin{align*}
&(1+\varepsilon)(1+\delta)\Pr\{\hat{Y}=1|Y=1, A=0\} \\
&= (1+\varepsilon)(1+\delta)\Pr\{\hat{Y}=1, \hat{X}=1|Y=1, A=0\}\\ 
&= (1+\varepsilon)\Pr\{\hat{X}=1|Y=1, A=0\}(1+\delta)\Pr\{\hat{Y}=1|\hat{X}=1, Y=1, A=0\}\\
&\leq \Pr\{\hat{X}=1|Y=1, A=1\}\Pr\{\hat{Y}=1|\hat{X}=1, Y=1, A=1\}\\
&= \Pr\{\hat{Y}=1, \hat{X}=1|Y=1, A=1\}\\
&= \Pr\{\hat{Y}=1|Y=1, A=1\}
\end{align*}
Therefore, $$(1+\varepsilon)(1+\delta)\Pr\{\hat{Y}=1|Y=1, A=0\} \leq \Pr\{\hat{Y}|Y=1, A=1\},$$ so the pipeline is $(1+\varepsilon)(1+\delta)=(1+\varepsilon + \delta + o(\varepsilon+\delta))$-equal opportunity fair and hence fairness is multiplicative over each stage under the above assumptions.

\subsubsection*{A Toy Example}
To provide insight into how a $(1+\varepsilon)$-equal opportunity decision at different stages of the pipeline affect outcomes, we give an example using the two-stage  hiring model pipeline. 

Suppose a company wishes to interview 20 people, and hire 2 of those 20. Assume 100 applicants apply, with 90 from a majority group and 10 from a minority group. Also assume the proportion of applicants qualified for the job are equal for both groups. For this simple example, we make the strong assumption that we have very good algorithms that choose only people qualified for the job, and that there are enough qualified applicants for each scenario.

Define the interview, the first stage, as a $(1+\varepsilon)$-equal opportunity decision, and the hiring, the second stage, as a $(1+\delta)$ decision. Using the definitions from the above section with strict equality, our decisions satisfy:

\begin{enumerate}
\item $(1+\varepsilon)\Pr\{\hat{X}=1|Y=1, A=0\} = \Pr\{\hat{X}=1|Y=1, A=1\}$
\item $(1+\delta)\Pr\{\hat{Y}=1|\hat{X}=1, Y=1, A=0\}= \Pr\{\hat{Y}=1|\hat{X}=1, Y=1, A=1\}$
\end{enumerate}

Table 1 and Figure 1 provide a numerical table and visual of four scenarios. Case one and two present a situation where a perceived bias is accounted for at different stages in the pipeline. Case three presents a scenario where an attempt to fix a bias is implemented at stage one, but is counterbalanced at stage two, and case four presents the reverse circumstance.

\begin{table}[h]
\centering
\caption{Four Cases}
    \begin{tabular}{p{0.17\linewidth}p{0.17\linewidth}p{0.17\linewidth}p{0.17\linewidth}p{0.17\linewidth}}
    \toprule
    Case: $\varepsilon$, $\delta$ & Expected majority \qquad interviewed & Expected \quad minority \qquad interviewed & Expected majority hired & Expected minority hired  \\ 
     \midrule
   1: 2, 0 & 15 & 5 & 1.5 & 0.5 \\
    \midrule
  2: 0, 2 & 18 & 2 & 1.5 & 0.5  \\ 
    \midrule
   3: 2, -.666 & 15 & 5 & 1.8 & 0.2  \\ 
    \midrule
  4: -.666, 2 & 19.28 & .71 & 1.8 & 0.2  \\ 
    \bottomrule
    \end{tabular}
    \label{Tab:1}
\end{table}

\begin{figure}[h]
  \centering
  \includegraphics[width=.5\textwidth]{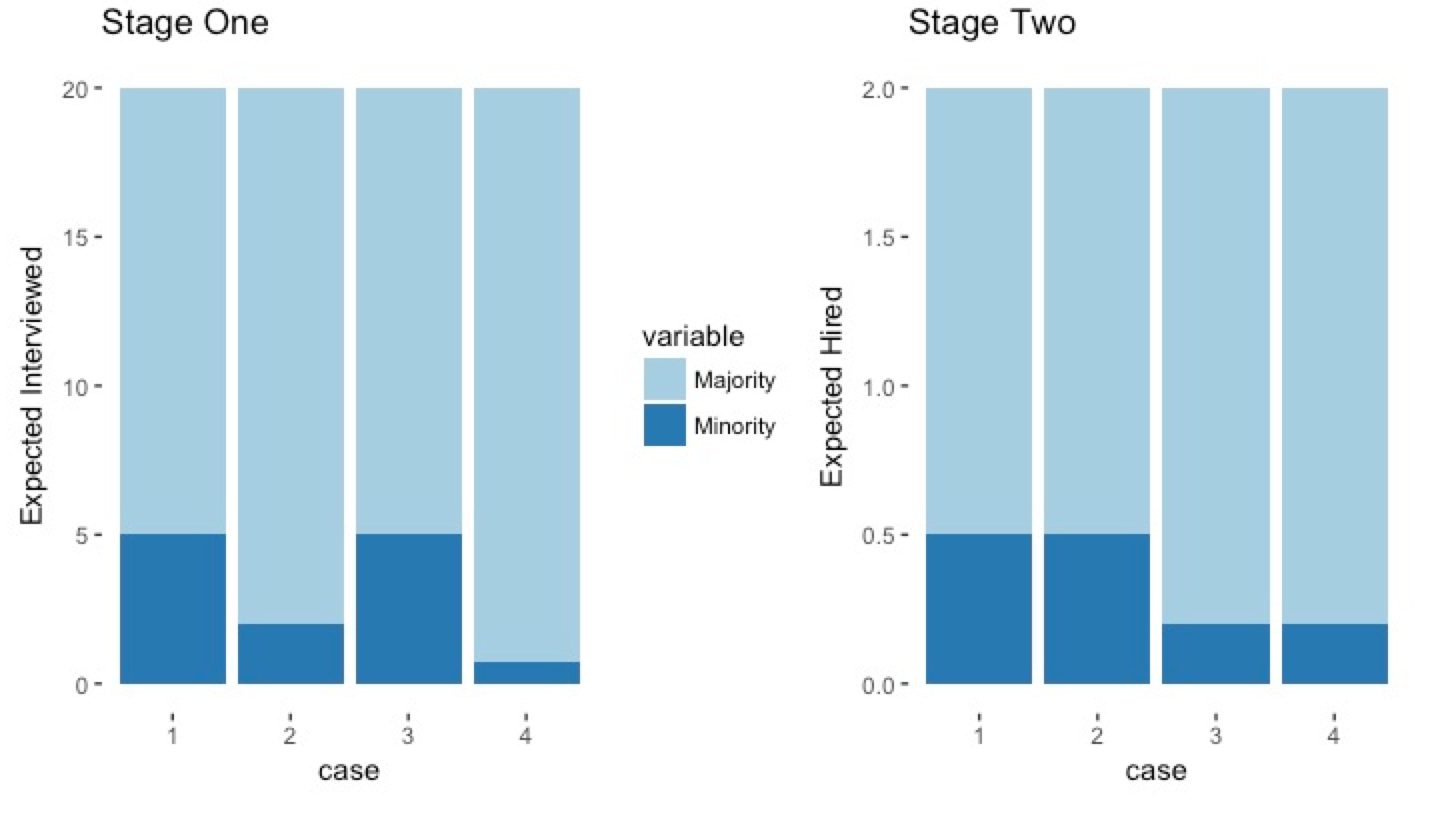}
  \caption{}
  \label{fig:f1}
\end{figure}

One observation of note is that if one wishes to implement a $(1+\varepsilon)$-decision to fix a perceived bias, the final outcome is independent of the stage at which the $(1+\varepsilon)$-decision is made. Simulation shows that the variance of the final decision is also independent of the stage at which a decision is implemented. 

This may be important for future policy, as giving preference to a minority group during the interview is perhaps more publicly acceptable than giving higher preference to minorities during hiring. Additionally, instead of giving preference to the minority group to receive more interviews from the current unbalanced applicant pool, the same outcome can be achieved by recruiting more minority applicants. 

\subsection{Where Difficulties Lie}
Notice that, above we assume that fairness in the first stage of the pipeline to mean $(1+\varepsilon)\Pr\{\hat{X}=1|Y=1, A=0\} \leq \Pr\{\hat{X}=1|Y=1, A=1\}$ and fairness in the second stage to mean $(1+\delta)\Pr\{\hat{Y}=1|\hat{X}=1, Y=1, A=0\}\leq \Pr\{\hat{Y}=1|\hat{X}=1, Y=1, A=1\}$. Fairness in stage two fits in the framework of equal opportunity since it's a statement about an applicant getting hired given that the applicant is hiring-qualified and made it successfully through the first stage of the pipeline. Unfortunately, the first stage ``fairness'' assumption is a bit troublesome because it requires the first stage to make a decision based on the quality measured in the last stage. In a real world application, the first stage may be controlled by different mechanisms or goals than the last stage.  In the context of a pipeline process where each portion is controlled by the same organization (or perhaps the portions are controlled by two sub-entities of the one organization), these assumptions make sense. However, there are many scenarios where this assumption will not be met. 

For an example, we return to the two-stage pipeline hiring example where the first stage of the pipeline determines who gets an interview and the second stage determines who gets hired. The interview stage may only care about someone's resume to determine if they should be granted an interview. However, it's not hard to imagine a case where a candidate is hiring qualified ($Y=1$; they have the skills for the interview and job) but is not interview qualified ($X = 0$; their resume could be bad because they never received guidance on creating a good resume). Therefore, the issue seems to be with the specific ratio for $i=0,1$: $$\frac{\Pr\{ \hat{X}=1 | X=1, A=i\}}{\Pr\{ \hat{X}=1 | Y=1, A=i\}}.$$ Ideally, we want these probabilities to be as close as possible so that we can decouple the pipeline. If not, being fair in the interview stage only based on interview qualifications and being fair in the hiring stage may not result in a pipeline that is fair.



\section{Conclusion}
In this work, we formalized the notion of a compound decision process called a pipeline, which is ubiquitous in domains like hiring, criminal justice, and finance. A pipeline decouples the final decision into intermediate decisions, which is important since although each decision affects the final outcome, different processes with different goals may be in charge of each intermediate stage. Decoupling allows us to see how fairness in each stage contributes to the fairness in the final decision. 

We also modified the definition of equal opportunity to allow boosting of the minority class and showed under what assumptions of fairness on the intermediate stages of the pipeline result in an overall fair pipeline according to this definition. In this case, the fairness from each stage of the pipeline is in some sense independent so that the entire pipeline has fairness factor given by the product. On the other hand, we would like to point out if the first stage is unfair, then the second stage cannot necessarily rectify the situation.  For example, if the first stage grants interviews to just one or two from a minority class, then the second stage is limited to hiring both of them, which will not result in overall fairness.  Therefore, it is important to get the first stage right.

\subsection{Future Work}\label{futurework}
We hope that our work highlights the need and sets the stage to understand compound decision making. We now give many directions for future research. 
\subsubsection*{Stability}
In a straight pipeline, will a small amount of unfairness or bias in the beginning turn into a large amount of unfairness at the end?
\subsubsection*{Bias as pipeline stage}  In our main hiring example above, we composed two remedies to implicit bias.  Alternatively, bias might be analyzed as a pipeline stage with parameter $\epsilon$ of sign opposite to the corresponding $\epsilon$ in the remedy stage.
\subsubsection*{Transparency} How transparent can a pipeline be? One method to test is whether measuring transparency by qualitative input influence \cite{reviewer} is cumulative, and how it differs by the type of pipeline. Does one need information at each stage, or is it sufficient to have good information of only the final decision?
\subsubsection*{Variance and small pools}
We can ask that the \textbf{expected} rate of positive decisions for a subclass be (approximately) proportional to that class's presence in the population.  But, as with reliable randomized algorithms, we really want the outcomes to concentrate at (or near) the expectation with high probability.  If a protected class is tiny, then small aberrations may cause an outcome far from the mean, with relatively large probability.  In the context of pipelines, after asking whether means are preserved through pipelines, we can ask whether concentration is preserved.  As in the case with randomized algorithms, it is sometimes appropriate to distort the mean to preserve concentration. 
\subsubsection*{Feedback loops}
More generally, some situations involve many interacting decisions, possibly with feedback loops.  Under what circumstances can each decision be made autonomously, with some guarantee that the system will converge to meet some overall guarantee of fairness?  For example, early in the days of long-distance running, just after women were allowed to enter major marathons without restriction, fewer women than men chose to do so at the elite level.  Some race organizers---on the assumption that the sport should attract women and men equally---offered an equal \textbf{total} purse (say, for the top ten spots) for each of the men's and women's races and, since fewer women entered, those who did chased a larger expected \textbf{individual} payout.  Policies like these are designed to lure more women elites the following year.
What if the process is decelerated, say, by offering a purse proportional to the square root of the participation rate, e.g., with initial participation rates of $r_0:r_1$ given by 10:90, the purse is proportional to $\sqrt{r}$, so $\sqrt{.1}$:$\sqrt{.9}$, which is 25:75?  What if the process is accelerated, by offering a purse proportional to $1/r$, so $1/.1:1/.9$, about 9:1, reversing the participation ratio?  Note that, for these types of incentives, Teither class needs to be marked as protected, which may be desirable, though the goal of equal participation must be assumed.  Typically (but depending on the strength of the economic signal to incentivize future runners), $\sqrt{r}$ has an attractive fixed point at 50:50 and $1/r$ has a repulsive fixed point at 50:50.  Under appropriate assumptions, we can pose a purely mathematical question: What is the least  function $g(r)$ (or give a bound on $g$) that leads to an attractive fixed point?
By analogy with cryptography, can such systems (with or without loops) be tolerant to a bounded number of unfair players?  Or bounded "total amount of unfairness," distributed among all the players and not otherwise quantified or understood?
\subsubsection*{Definitions of fairness}
To return to the elite runners example, we may need refined definitions of fairness, say, "interim fairness," that captures increased year-over-year participation of an underrepresented class, and calls the improvement "interim fair" even if the system has not yet converged to a fair state.  As for hiring, suppose a University department only hires faculty from a pool of recent PhDs, which is unbalanced.  If the hiring process selects underrepresented faculty at a far higher rate than their presence in the PhD pool, that should be deemed "interim fair" for some purposes.  




\subsubsection*{Different notions of fairness}
Furthermore, we would like to understand how different definitions of fairness other than $(1+\varepsilon)$-equal opportunity propagate through a pipeline.

\section*{Acknowledgement}
This material is based upon work supported by the National Science Foundation Graduate Research Fellowship under Grant No. DGE 1256260 as well as by the NSF grants CCF-1161233 and IIS-1633724.



\bibliographystyle{ACM-Reference-Format}
\bibliography{bibliography} 

\end{document}